\documentclass[lettersize,journal]{IEEEtran}
\usepackage{amsmath,amsfonts}
\usepackage{algorithmic}
\usepackage{algorithm}
\usepackage{array}
\usepackage[caption=false,font=normalsize,labelfont=sf,textfont=sf]{subfig}
\usepackage{textcomp}
\usepackage{stfloats}
\usepackage{url}
\usepackage{verbatim}
\usepackage{graphicx}
\usepackage{cite}

\usepackage{booktabs}
\usepackage{multirow}
\usepackage{caption} 
\usepackage{subcaption} 
\usepackage{multirow, multicol}
\usepackage{color}
 \usepackage[colorlinks]{hyperref}

\usepackage{hyperref}

\begin{document}

\title{VM-UNet: Vision Mamba UNet for Medical Image Segmentation}
\author{Jiacheng Ruan, Jincheng Li, and Suncheng Xiang,~\IEEEmembership{Member,~IEEE}
\thanks{Corresponding author: Suncheng Xiang.}
\thanks{Jiacheng Ruan is with the School of Electronic Information and Electrical Engineering, Shanghai Jiao Tong University, No 800, Dongchuan Road, Minhang District, Shanghai, China, 200240  (email: jackchenruan@sjtu.edu.cn).}
\thanks{Jincheng Li, Suncheng Xiang are with the School of Biomedical Engineering, Shanghai Jiao Tong University, No 800, Dongchuan Road, Minhang District, Shanghai, China, 200240 (email: ljc2293@sjtu.edu.cn; xiangsuncheng17@sjtu.edu.cn).}}


\markboth{Journal of \LaTeX\ Class Files,~Vol.~14, No.~8, August~2021}%
{Shell \MakeLowercase{\textit{et al.}}: A Sample Article Using IEEEtran.cls for IEEE Journals}


\maketitle

\begin{abstract}
In the realm of medical image segmentation, both CNN-based and Transformer-based models have been extensively explored. However, CNNs exhibit limitations in long-range modeling capabilities, whereas Transformers are hampered by their quadratic computational complexity. Recently, State Space Models (SSMs), exemplified by Mamba, have emerged as a promising approach. They not only excel in modeling long-range interactions but also maintain a linear computational complexity. In this paper, leveraging state space models, we propose a U-shape architecture model for medical image segmentation, named Vision Mamba UNet (VM-UNet). Specifically, the Visual State Space (VSS) block is introduced as the foundation block to capture extensive contextual information, and an asymmetrical encoder-decoder structure is constructed with fewer convolution layers to save calculation cost. We conduct comprehensive experiments on the ISIC17, ISIC18, and Synapse datasets, and the results indicate that VM-UNet performs competitively in medical image segmentation tasks. To our best knowledge, this is the first medical image segmentation model constructed based on the pure SSM-based model. We aim to establish a baseline and provide valuable insights for the future development of more efficient and effective SSM-based segmentation systems. Our code is available at \href{https://github.com/JCruan519/VM-UNet}{https://github.com/JCruan519/VM-UNet}.
\end{abstract}


\section{Introduction}
\begin{figure}[!t]
    \centering
    \includegraphics[width=0.50\textwidth]{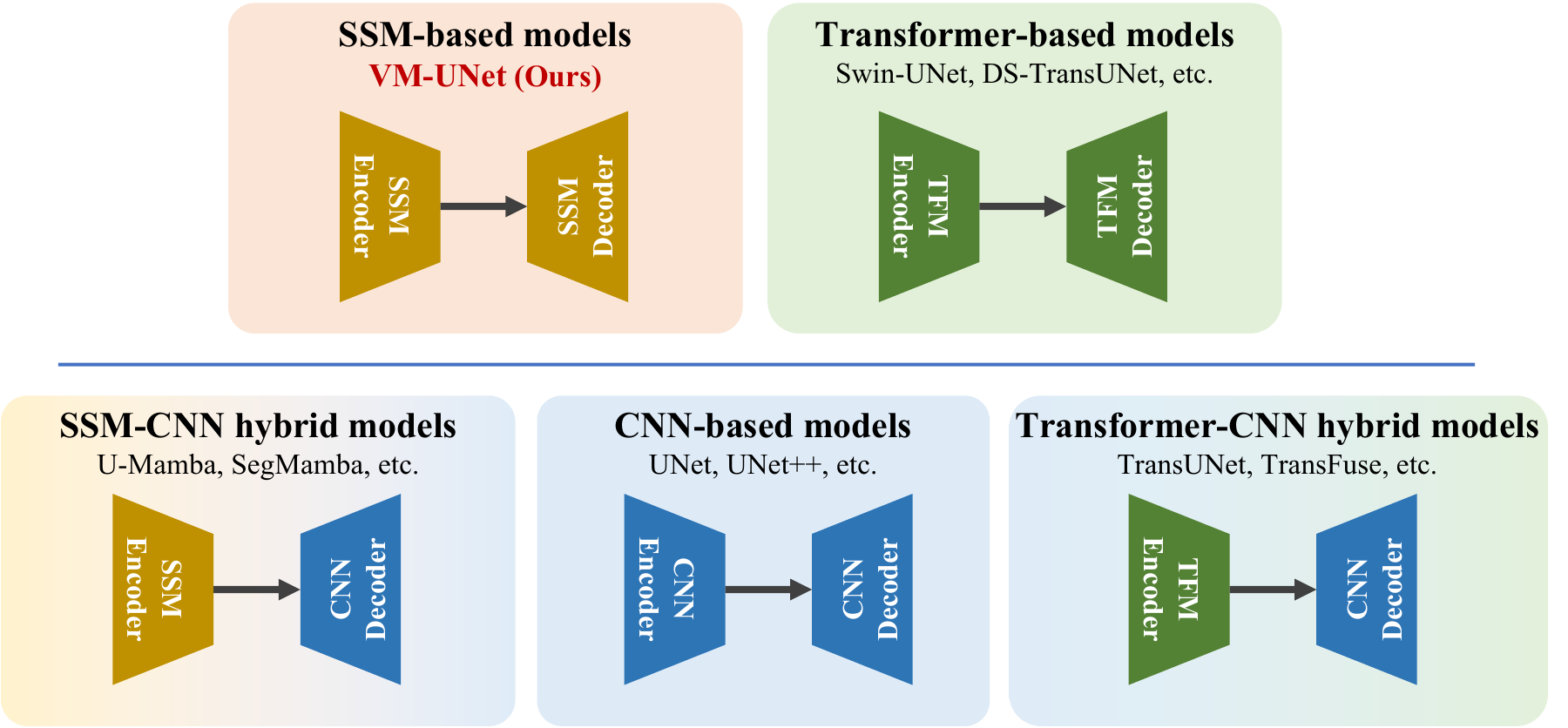}
    \caption{A brief introduction to the development of U-shape architecture in the field of medical image segmentation.}
    \label{fig:head}
\end{figure}

\IEEEPARstart{A}{utomated} medical image segmentation techniques assist physicians in faster pathological diagnosis, thereby improving the efficiency of patient care. Recently, CNN-based and Transformer-based models have demonstrated remarkable performance in various visual tasks, particularly in medical image segmentation. UNet \cite{unet}, as a representative of CNN-based models, is known for its simplicity of structure and strong scalability, and many subsequent improvements are based on this U-shaped architecture \cite{3DUNet,unet++,malunet,egeunet,mewunet,li2023erdunet,zhou2024uncertainty}. TransUnet \cite{transunet}, a pioneer among Transformer-based models, is the first to employ Vision Transformer (ViT) \cite{vit} for feature extraction during the encoding phase and utilizes CNN in the decoding phase, demonstrating the significant capability for global information acquisition. Subsequently, TransFuse \cite{transfuse} incorporates a parallel architecture of ViT and CNN, capturing both local and global features simultaneously. Furthermore, Swin-UNet \cite{swinunet} combines Swin Transformer \cite{swintrm} with the U-shaped architecture, introducing a pure Transformer-based U-shaped model for the first time, more detailed development of U-shape architecture is illustrated in Figure~\ref{fig:head}.

Nevertheless, both CNN-based models and Transformer-based models have inherent limitations~\cite{liu2023joint}. CNN-based models are constrained by their local receptive field, considerably hindering their ability to capture long-range information. This often leads to the extraction of inadequate features, resulting in suboptimal segmentation outcomes. Although Transformer-based models demonstrate superior performance for global modeling, the self-attention mechanism demands quadratic complexity in terms of image sizes, leading to a high computational burden \cite{transformer,vit,mu2023learning}, particularly for tasks requiring dense predictions like medical image segmentation. The current shortcomings in these models compel us to develop a novel architecture for medical image segmentation, capable of capturing strong long-range information and maintaining linear computational complexity.

Recently, State Space Models (SSMs) have attracted considerable interest among researchers. Building on the foundation of classical SSM \cite{ssm1960} research, the modern SSMs (e.g., Mamba \cite{mamba}) not only establish long-distance dependencies but also exhibit linear complexity with respect to input size. Additionally, SSM-based models have received substantial research across many fields, including language understanding \cite{ssm-s4,mamba}, general vision \cite{VisionMamba,vmamba}, etc. Particularly, U-Mamba \cite{Umamba} has recently introduced a novel SSM-CNN hybrid model, marking its first application in medical image segmentation tasks. SegMamba \cite{segmamba} incorporates SSM in the encoder part, while still using CNN in the decoder part, suggesting a SSM-CNN hybrid model for 3D brain tumor segmentation tasks. Although aforementioned works have utilized SSM for medical image segmentation tasks, the performance of the pure SSM-based model has yet to be explored.

Influenced by the success of VMamba \cite{vmamba} in image classification tasks, this paper introduces the Vision Mamba UNet (VM-UNet) for the first time, a pure SSM-based model designed to showcase the potential in medical image segmentation tasks. Specifically, VM-UNet is composed of three main parts: the encoder, the decoder, and the skip connection. The encoder consists of VSS blocks from VMamba for feature extraction, along with patch merging operations for downsampling. Conversely, the decoder comprises VSS blocks and patch expanding operations to restore the size of the segmentation results. For the skip connection component, to highlight the segmentation performance of the most original pure SSM-based model, we adopt the simplest form of additive operation.

Comprehensive experiments are conducted on organ segmentation and skin lesion segmentation tasks to demonstrate the potential of pure SSM-based models in medical image segmentation. Specifically, we conduct extensive experiments on the Synapse \cite{synapse}, ISIC17 \cite{isic17}, and ISIC18 \cite{isic18} datasets, the results of which indicate that VM-UNet can achieve competitive performance. Moreover, it is important to note that VM-UNet represents the most basic form of a pure SSM-based segmentation model, as it does not include any specially designed modules.

The main contributions of this paper can be summarized as follows:
 \begin{itemize}
\item We propose VM-UNet, marking the first occasion of exploring the potential applications of purely SSM-based models in medical image segmentation.
\item Comprehensive experiments are conducted on three datasets, with results indicating that VM-UNet exhibits considerable competitiveness.
\item We establish a baseline for pure SSM-based models in medical image segmentation tasks, providing valuable insights that pave the way for the development of more efficient and effective SSM-based methods in medical image.
\end{itemize}

In the rest of the paper, we first review some related works of CNN-based methods, Transformer-based methods and SM-based models for medical image segmentation in Section \ref{sec2}. Then in Section \ref{sec3}, we give more details about the architecture and learning procedure of the proposed VM-UNet method. Extensive evaluations compared with state-of-the-art methods and comprehensive analyses of the proposed approach are elaborated in Section \ref{sec4}. Conclusion and Future Works are given in Section \ref{sec5}.

\section{Related Works}
\label{sec2}

In this section, we give a brief review of the related works on common medical image segmentation methods. The core idea of
these existing methods is to efficiently model contextual information. These methods
can be roughly divided into CNN-based approaches, Transformer-based approaches and SSM-based approaches.

\subsection{CNN-based Models}

Traditional medical image segmentation methods are commonly based on machine learning approaches \cite{MLMIS,sun2003semiautomatic}. With the advancement of convolutional neural networks, UNet \cite{unet}, as an end-to-end segmentation model, has achieved commendable performance in medical image segmentation. Furthermore, due to UNet's simple structure, strong scalability, and effective segmentation capabilities, many subsequent works have been based on modifications of this U-shaped architecture. For instance, UNet++ \cite{unet++} replaces the simple skip connections in UNet with dense connections to enhance feature representation and compensate for information loss caused by sampling. Attention-UNet \cite{attentionunet} employs attention gates to control the importance of features, thereby enabling the model to focus more on target areas. MALUNet \cite{malunet} introduces four efficient attention modules, significantly reducing the parameter count and computational load of UNet, while achieving superior segmentation performance. For medical image segmentation, understanding global image information is crucial; however, CNN-based models inherently struggle with modeling long-range dependencies.

\subsection{Transformer-based Models}

Influenced by the popularity of ViT \cite{vit} in the general vision domain, many works have introduced Transformer-based models into medical image segmentation tasks to enhance the capability of the models in modeling long-range dependencies. For instance, TransFuse \cite{transfuse} employs a parallel encoder structure, using both CNN and ViT to process local and global features. Swin-UNet \cite{swinunet} combines Swin Transformer \cite{swintrm} with UNet, presenting the first purely Transformer-based medical image segmentation model. DS-TransUNet \cite{dstransunet} inputs patches of varying sizes and employs two parallel Swin Transformers for encoding, aiming to capture multi-scale feature information. MEW-UNet \cite{mewunet} introduces frequency domain operations into the Transformer block, proposing the Multi-axis External Weights block to capture richer frequency domain features. However, Transformer-based models necessitate quadratic complexity with respect to input sizes, resulting in a substantial computational burden.

\subsection{SSM-based Models}

SSM-based models, such as HiPPO \cite{hippo}, LSSL \cite{LSSL}, and others, which are derived from linear state-space equations in control theory, are primarily employed for sequence data modeling tasks. The Structured State-Space Sequence (S4) model, introduced in \cite{ssm-s4}, facilitates long-distance modeling with only linear scaling in sequence length. Recently, Mamba \cite{mamba} has emerged as a novel alternative to CNNs and Transformers. It extends the S4 framework by integrating a data-dependent SSM layer, yielding competitive results in natural language processing tasks and further propelling the evolution of SSM-based models. Inspired by Mamba's success, Vision Mamba \cite{VisionMamba} and VMamba \cite{vmamba} have implemented this SSM-based model with a data-dependent SSM layer in the general vision domain, achieving notable performance. Additionally, U-Mamba \cite{Umamba} and SegMamba \cite{segmamba}, as hybrid SSM-CNN models, have begun to demonstrate the potential of SSM-based models in medical image segmentation. However, the capabilities of purely SSM-based models in segmentation have not been fully explored. Consequently, this paper presents the first purely SSM-based model named VM-UNet, which can be regarded as the foundation block to capture extensive contextual information for medical image segmentation.

\section{Methods}
\label{sec3}

\subsection{Preliminaries}

In modern SSM-based models, i.e., Structured State Space Sequence Models (S4) and Mamba, both rely on a classical continuous system that maps a one-dimensional input function or sequence, denoted as $x(t) \in \mathcal{R}$, through intermediate implicit states $h(t) \in \mathcal{R}^N$ to an output $y(t) \in \mathcal{R}$. The aforementioned process can be represented as a linear Ordinary Differential Equation (ODE):

\begin{equation}
\begin{aligned}
    &h^{\prime}(t) = \mathbf{A}h(t) + \mathbf{B}x(t) \\
    &y(t) = \mathbf{C}h(t)
\end{aligned}
\end{equation} where $\mathbf{A} \in \mathcal{R}^{N \times N}$ represents the state matrix, while $\mathbf{B} \in  \mathcal{R}^{N \times 1}$ and $\mathbf{C} \in  \mathcal{R}^{N \times 1}$ denote the projection parameters.

S4 and Mamba discretize this continuous system to make it more suitable for deep learning scenarios. Specifically, they introduce a timescale parameter $\mathbf{\Delta}$ and transform $\mathbf{A}$ and $\mathbf{B}$ into discrete parameters $\overline{\mathbf{A}}$ and $\overline{\mathbf{B}}$ using a fixed discretization rule. Typically, the zero-order hold (ZOH) is employed as the discretization rule and can be defined as follows:

\begin{equation}
\begin{aligned}
    &\overline{\mathbf{A}} = \textup{exp}(\mathbf{\Delta} \mathbf{A}) \\
    &\overline{\mathbf{B}} = (\mathbf{\Delta} \mathbf{A})^{-1}(\textup{exp}(\mathbf{\Delta} \mathbf{A}) - \mathbf{I})\cdot\mathbf{\Delta} \mathbf{B}
\end{aligned}
\end{equation}

After discretization, SSM-based models can be computed in two ways: linear recurrence or global convolution, defined as equations \ref{eq:linear_recurrence} and \ref{eq:global_convolution}, respectively.

\begin{equation}
\begin{aligned}
    &h^{\prime}(t) = \overline{\mathbf{A}}h(t) + \overline{\mathbf{B}}x(t) \\
    &y(t) = \mathbf{C}h(t)
\end{aligned}
\label{eq:linear_recurrence}
\end{equation}

\begin{equation}
\begin{aligned}
    &\overline{K} = (\mathbf{C}\overline{\mathbf{B}}, \mathbf{C}\overline{\mathbf{AB}}, \ldots, \mathbf{C}\overline{\mathbf{A}}^{L-1}\overline{\mathbf{B}}) \\
    &y = x * \overline{\mathbf{K}}
\end{aligned}
\label{eq:global_convolution}
\end{equation}where $\overline{\mathbf{K}} \in \mathcal{R}^{L}$ represents a structured convolutional kernel, and $L$ denotes the length of the input sequence $x$.

\begin{figure}[!t]
    \centering
    \includegraphics[width=0.50\textwidth]{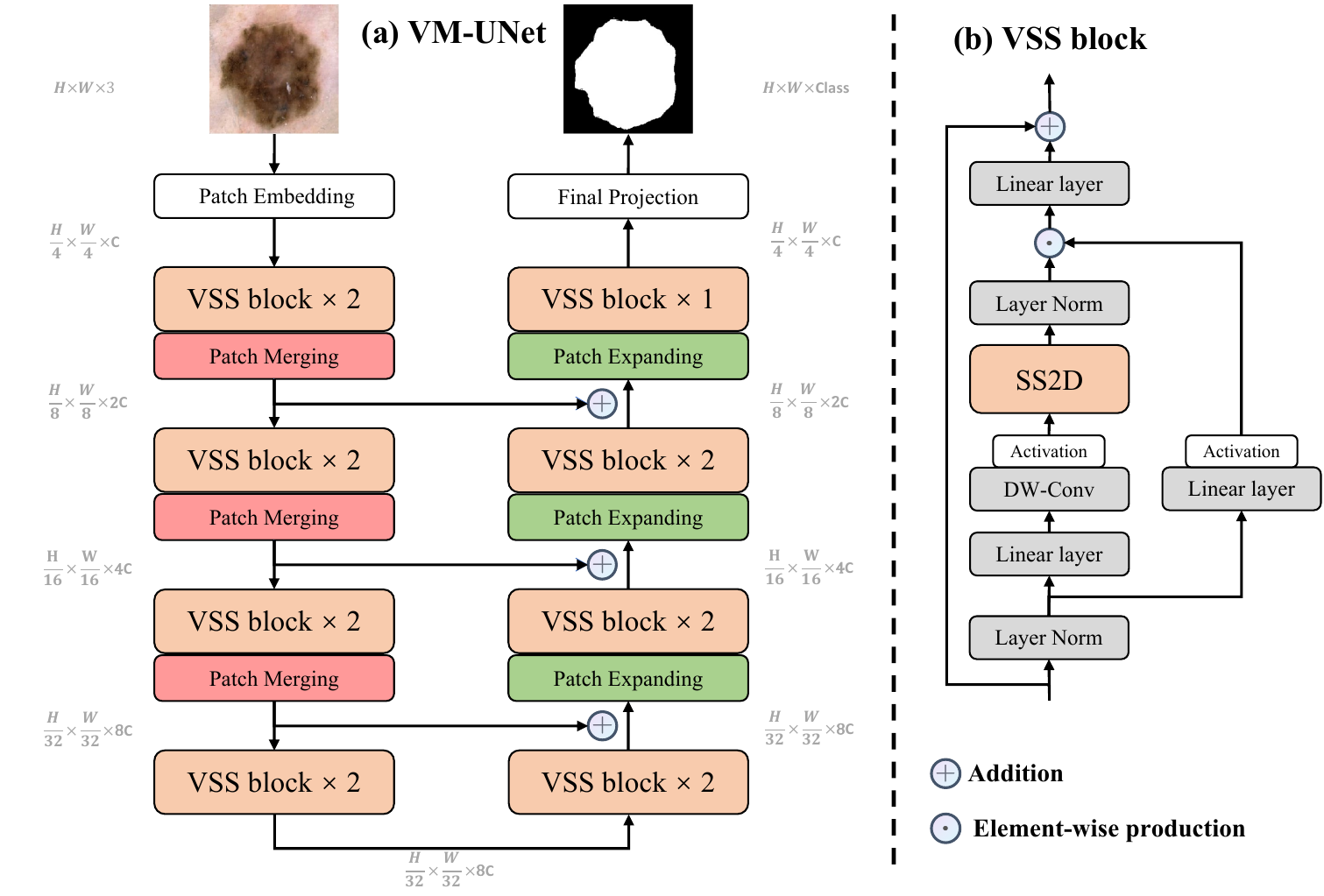}
    \caption{(a) The overall architecture of VM-UNet. (b) VSS block is the main construction block of VM-UNet, and SS2D is the core operation in VSS block.}
    \label{fig:main}
\end{figure}

In this section, we initially introduce the overall structure of VM-UNet. Subsequently, we elaborate on the core component, the VSS block. Finally, we describe the loss function utilized during the training process.

\subsection{Vision Mamba UNet (VM-UNet)}

As depicted in Figure \ref{fig:main} (a), the overall architecture of VM-UNet is presented. Specifically, VM-UNet comprises a Patch Embedding layer, an encoder, a decoder, a Final Projection layer, and skip connections. Unlike previous methods \cite{swinunet}, we have not adopted a symmetrical structure but instead utilized an asymmetric design.

The Patch Embedding layer divides the input image $x \in \mathcal{R}^{H \times W \times 3}$ into non-overlapping patches of size $4\times4$, subsequently mapping the dimensions of the image to $C$, with $C$ defaulting to 96. This process results in the embedded image $x^{\prime} \in \mathcal{R}^{\frac{H}{4} \times \frac{W}{4} \times C}$. Finally, we normalize $x^{\prime}$ using Layer Normalization \cite{layernorm} before feeding it into the encoder for feature extraction. The encoder is composed of four stages, with a patch merging operation applied at the end of first three stages to reduce the height and width of the input features while increasing the number of channels. We employ [2, 2, 2, 2] VSS blocks across four stages, with the channel counts for each stage being [C, 2C, 4C, 8C].

Similarly, the decoder is organized into four stages. At the beginning of last three stages, a patch expanding operation is utilized to decrease the number of feature channels and increase the height and width. Across the four stages, we utilize [2, 2, 2, 1] VSS blocks, with the channel counts for each stage being [8C, 4C, 2C, C]. Following the decoder, a Final Projection layer is employed to restore the size of the features to match the segmentation target. Specifically, a 4-times upsampling is conducted via patch expanding to recover the height and width of the features, followed by a projection layer to restore the number of channels.

For the skip connections, a straightforward addition operation is adopted without bells and whistles, thereby not introducing any additional parameters.

\subsection{VSS Block}

The VSS block derived from VMamaba \cite{vmamba} is the core module of VM-UNet, as depicted in Figure \ref{fig:main} (b). After undergoing Layer Normalization, the input is split into two branches. In the first branch, the input passes through a linear layer followed by an activation function. In the second branch, the input undergoes processing through a linear layer, depthwise separable convolution, and an activation function, before being fed into the 2D-Selective-Scan (SS2D) module for further feature extraction. Subsequently, the features are normalized using Layer Normalization, and then an element-wise production is performed with the output from the first branch to merge the two pathways. Finally, the features are mixed using a linear layer, and this outcome is combined with a residual connection to form the VSS block's output. In this paper, SiLU \cite{silu} is employed as the activation function by default.

\begin{figure*}[!t]
    \centering
     \includegraphics[width=1.00\textwidth]{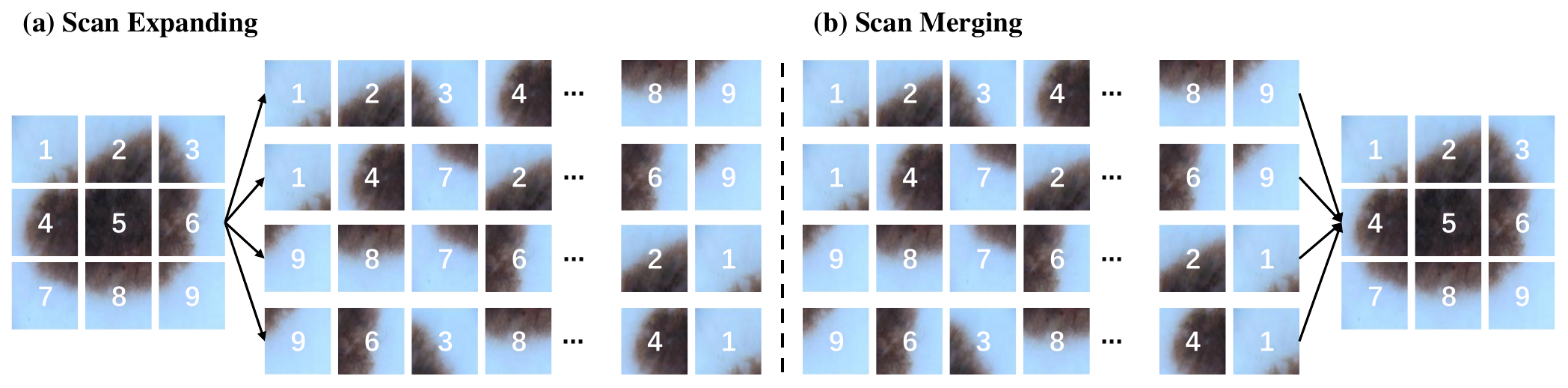}
    \caption{(a) The scan expanding operation in SS2D. (b) The scan merging operation in SS2D.}
    \label{fig:scan}
\end{figure*}

The SS2D consists of three components: a scan expanding operation, an S6 block, and a scan merging operation. As shown in Figure \ref{fig:scan}(a), the scan expanding operation unfolds the input image along four different directions (top-left to bottom-right, bottom-right to top-left, top-right to bottom-left, and bottom-left to top-right) into sequences. These sequences are then processed by the S6 block for feature extraction, ensuring that information from various directions is thoroughly scanned, thus capturing diverse features. Subsequently, as illustrated in Figure \ref{fig:scan}(b), the scan merging operation sums and merges the sequences from the four directions, restoring the output image to the same size as the input. The S6 block, derived from Mamba \cite{mamba}, introduces a selective mechanism on top of S4 \cite{ssm-s4} by adjusting the SSM's parameters based on the input. This enables the model to distinguish and retain pertinent information while filtering out the irrelevant. The pseudo-code for the S6 block is presented in Algorithm \ref{alg:s6}.

\subsection{Loss Function}

The introduction of VM-UNet is aimed at validating the application potential of pure SSM-based models in medical image segmentation tasks. Consequently, we exclusively utilize the most fundamental Binary Cross-Entropy and Dice loss (BceDice loss) and Cross-Entropy and Dice loss (CeDice loss) as the loss functions for binary and multi-class segmentation tasks, respectively, as denoted by Equation \ref{eq:bcediceloss} and  Equation \ref{eq:cediceloss}.

\begin{equation}
    L_{\text{BceDice}} = \lambda_1 * L_{\text{Bce}} + \lambda_2 * L_{\text{Dice}}
    \label{eq:bcediceloss}
\end{equation}

\begin{equation}
    L_{\text{CeDice}} = \lambda_3 * L_{\text{Ce}} + \lambda_4 * L_{\text{Dice}}
    \label{eq:cediceloss}
\end{equation}

\begin{equation}
\left\{
\begin{aligned}
    L_{\text{Bce}} &= -\frac{1}{N} \sum_{i=1}^{N} \left[ y_i \log(\hat{y}_i) + (1 - y_i) \log(1 - \hat{y}_i) \right] \\
    L_{\text{Ce}} &= -\frac{1}{N} \sum_{i=1}^{N} \sum_{c=1}^{C} y_{i,c} \log(\hat{y}_{i,c}) \\
    L_{\text{Dice}} &= 1 - \frac{2|X \cap Y|}{|X|+|Y|}
\end{aligned}
\right.
\end{equation} 
where \(N\) denotes the total number of samples, and \(C\) represents the total number of categories. \(y_i, \hat{y}_i\) respectively signify the true label and prediction label. \(y_{i,c}\) is an indicator that equals 1 if sample \(i\) belongs to category \(c\), and 0 otherwise. \(\hat{y}_{i,c}\) is the probability that the model predicts sample \(i\) as belonging to category \(c\). \(|X|\) and \(|Y|\) represent the ground truth and prediction, respectively. \(\lambda_1, \lambda_2,  \lambda_3, \lambda_4\) control the relative importance of different loss terms in Equation~\ref{eq:bcediceloss} and Equation~\ref{eq:cediceloss}, which are empirically set to 1 by default.

To sum up, we propose a simple but effective training framework named VM-UNet, which can explore the potential applications of purely SSM-based models in medical image segmentation. In addition, we establish a baseline for pure SSM-based models in medical image segmentation tasks, providing valuable insights that pave the way for the development of more efficient and effective SSM-based methods in medical images. To our best knowledge, this is the first medical image segmentation model constructed based on the pure SSM-based model. We hope this work can shed light for some potential tasks in medical community to move forward.

\begin{algorithm}[!t]
    \caption{Pseudo-code for S6 block in SS2D}
    \label{alg:s6}
    \renewcommand{\algorithmicensure}{\textbf{Output:}}
    \begin{algorithmic}[1]
        \renewcommand{\algorithmicrequire}{\textbf{Input:}}
        \REQUIRE the feature $x$ with shape [B, L, D] (batch size, token length, dimension)  
        \renewcommand{\algorithmicrequire}{\textbf{Params:}}
        \REQUIRE the nn.Parameter $\mathbf{A}$; the nn.Parameter $\mathbf{D}$
        \renewcommand{\algorithmicrequire}{\textbf{Operator:}}
        \REQUIRE Linear(.), the linear projection layer
        \ENSURE the feature $y$ with shape [B, L, D]    
        
        \STATE  $\mathbf{\Delta}, \mathbf{B}, \mathbf{C}$ = Linear($x$), Linear($x$), Linear($x$)
        \STATE  $\overline{\mathbf{A}} = \textup{exp}(\mathbf{\Delta} \mathbf{A})$
        \STATE  $\overline{\mathbf{B}} = (\mathbf{\Delta} \mathbf{A})^{-1}(\textup{exp}(\mathbf{\Delta} \mathbf{A}) - \mathbf{I})\cdot\mathbf{\Delta} \mathbf{B}$
        \STATE  $h_t = \overline{\mathbf{A}}h_{t-1} + \overline{\mathbf{B}}x_t$
        \STATE  $y_t = \mathbf{C}h_t + \mathbf{D}x_t$
        \STATE  $y = [y_1, y_2, \cdots, y_t, \cdots, y_L]$
        \RETURN $y$
    \end{algorithmic}
\end{algorithm}

\section{Experiments}
\label{sec4}

In this section, we conduct comprehensive experiments on VM-UNet for skin lesion and organ segmentation tasks. Specifically, we evaluate the performance of VM-UNet on medical image segmentation tasks on the ISIC17, ISIC18, and Synapse datasets.

\subsection{Datasets}

\textbf{ISIC17 and ISIC18 datasets:} The International Skin Imaging Collaboration 2017 and 2018 challenge datasets (ISIC17 and ISIC18) \cite{isic17,isic2017url,isic18,isic2018url} are two publicly available skin lesion segmentation datasets, containing 2,150 and 2,694 dermoscopy images with segmentation mask labels, respectively. Following the previous work \cite{malunet}, we split the datasets in a 7:3 ratio for use as training and test sets. Specifically, for the ISIC17 dataset, the training set consists of 1,500 images, and the test set consists of 650 images. For the ISIC18 dataset, the training set includes 1,886 images, while the test set contains 808 images. For these two datasets, we provide detailed evaluations on several metrics, including Mean Intersection over Union (mIoU), Dice Similarity Coefficient (DSC), Accuracy (Acc), Sensitivity (Sen), and Specificity (Spe).

\begin{table}[!t]
	\setlength\tabcolsep{1.3pt}
	\renewcommand\arraystretch{1.5}
	\caption{Comparative experimental results on the ISIC17 and ISIC18 dataset. (\textbf{Bold} indicates the best.)}
	\begin{center}
		\begin{tabular}{c|c|ccccc}
			\hline
			\textbf{Dataset} &\textbf{Model}          & \textbf{mIoU(\%)$\uparrow$}  & \textbf{DSC(\%)$\uparrow$}   & \textbf{Acc(\%)$\uparrow$}   & \textbf{Spe(\%)$\uparrow$}   & \textbf{Sen(\%)$\uparrow$}   \\ \hline
			\multirow{5}{*}{ISIC17} &UNet \cite{unet} &76.98 &86.99 &95.65 &97.43 & 86.82  \\
			&UTNetV2 \cite{utnetv2}                & 77.35          & 87.23          & 95.84          & 98.05          & 84.85          \\
			&TransFuse \cite{transfuse}              & 79.21          & 88.40          & 96.17          & 97.98          & 87.14 \\
            &MALUNet \cite{malunet} &78.78 &88.13 &96.18 &\textbf{98.47} &84.78 \\
			&\textbf{VM-UNet} &\textbf{80.23} &\textbf{89.03} &\textbf{96.29} &97.58 &\textbf{89.90}       \\ \hline\hline
			\multirow{8}{*}{ISIC18}&UNet \cite{unet}                    & 77.86          & 87.55          & 94.05          & \textbf{96.69}          & 85.86          \\
			&UNet++ \cite{unet++}                 & 78.31          & 87.83          & 94.02          & 95.75          & 88.65          \\
			&Att-UNet \cite{attentionunet}               & 78.43          & 87.91          & 94.13          & 96.23          & 87.60          \\
			&UTNetV2 \cite{utnetv2}                & 78.97          & 88.25          & 94.32          & 96.48          & 87.60          \\
			&SANet \cite{sanet}                  & 79.52          & 88.59          & 94.39          & 95.97          & 89.46          \\
			&TransFuse \cite{transfuse}              & 80.63          & 89.27          & 94.66          & 95.74          & \textbf{91.28} \\
            &MALUNet \cite{malunet} &80.25 &89.04 &94.62 &96.19 &89.74 \\
			&\textbf{VM-UNet} &\textbf{81.35} &\textbf{89.71} &\textbf{94.91}  &96.13 &91.12        \\ \hline
		\end{tabular}
		\label{tab:isic}
	\end{center}
\end{table}

\begin{table*}[!t]
	\setlength\tabcolsep{5.9pt}
	\renewcommand\arraystretch{1.5}
	\small
	\caption{Comparative experimental results on the Synapse dataset. DSC of every single class (Aorta, Gallbladder, Kidney(L), Kidney(R), Liver, Pancreas, Spleen, Stomach) is also reported. (\textbf{Bold} indicates the best.)}
	\begin{center}
		\begin{tabular}{c|cc|cccccccc}
			\hline
			\textbf{Model}          & \textbf{DSC$\uparrow$} & \textbf{HD95$\downarrow$} & \textbf{Aor.} & \textbf{Gal.} & \textbf{Kid.(L)} & \textbf{Kid.(R)} & \textbf{Liv.} & \textbf{Pan.} & \textbf{Spl.} & \textbf{Sto.} \\ \hline
			V-Net \cite{vnet}                   & 68.81            & -                 & 75.34          & 51.87                & 77.10              & 80.75     & 87.84          & 40.05             & 80.56           & 56.98            \\
   DARR \cite{DARR}  &69.77 &- &74.74 &53.77 &72.31 &73.24 &94.08 &54.18 &89.90 &45.96 \\
   R50 U-Net \cite{transunet} &74.68 &36.87 &87.47 &66.36 &80.60 &78.19 &93.74 &56.90 &85.87 &74.16 \\
			UNet \cite{unet}                  & 76.85            & 39.70             & 89.07 & \textbf{69.72}       & 77.77              & 68.60              & 93.43          & 53.98             & 86.67           & 75.58            \\
			Att-UNet \cite{attentionunet}                & 77.77            & 36.02             & \textbf{89.55}          & 68.88                & 77.98              & 71.11              & 93.57          & 58.04             & 87.30           & 75.75            \\
			TransUnet \cite{transunet}              & 77.48            & 31.69             & 87.23          & 63.13                & 81.87              & 77.02              & 94.08        & 55.86             & 85.08           & 75.62            \\
TransNorm \cite{transnorm}  &78.40 &30.25 &86.23 &65.10 &82.18 &78.63 &94.22 &55.34 &89.50 &76.01\\
Swin U-Net \cite{swinunet}  &79.13 &21.55 &85.47 &66.53 &83.28 &79.61 &\textbf{94.29} &56.58 &\textbf{90.66} &76.60 \\
TransDeepLab \cite{transdeeplab} &80.16 &21.25 &86.04 &69.16 &84.08 &79.88 &93.53 &\textbf{61.19} &89.00 &78.40 \\
			UCTransNet \cite{uctransnet}             & 78.23            & 26.75             & -              & -                    & -                  & -                  & -              & -                 & -               & -                \\
			MT-UNet  \cite{mtunet}               & 78.59            & 26.59             & 87.92          & 64.99                & 81.47              & 77.29              & 93.06          & 59.46    & 87.75           & 76.81           \\ 
			MEW-UNet \cite{mewunet} & 78.92   & \textbf{16.44}    & 86.68          & 65.32                & 82.87             & 80.02              & 93.63          & 58.36             & 90.19  & 74.26          \\ 
   \hline
   \textbf{VM-UNet} &\textbf{81.08}&19.21&86.40&69.41&\textbf{86.16}&\textbf{82.76}&94.17&58.80&89.51&\textbf{81.40} \\
   \hline 
		\end{tabular}
		\label{tab:synapse}
	\end{center}
\end{table*}

\textbf{Synapse multi-organ segmentation dataset (Synapse):} Synapse \cite{synapse,synapseurl} is a publicly available multi-organ segmentation dataset comprising 30 abdominal CT cases with 3,779 axial abdominal clinical CT images, including 8 types of abdominal organs (aorta, gallbladder, left kidney, right kidney, liver, pancreas, spleen, stomach). Following the setting of previous works \cite{transunet,swinunet}, 18 cases are used for training and 12 cases for testing. For this dataset, we report the Dice Similarity Coefficient (DSC) and the 95\% Hausdorff Distance (HD95).

\subsection{Implementation Details}
Following the prior works \cite{malunet,swinunet}, we resize the images in the ISIC17 and ISIC18 datasets to 256×256, and those in the Synapse dataset to 224×224. To prevent overfitting, data augmentation techniques, including random flip and random rotation, are employed. The BceDice loss function is utilized for the ISIC17 and ISIC18 datasets, while the CeDice loss function is adopted for the Synapse dataset. We set the batch size to 32 and employ AdamW \cite{adamw} optimizer with an initial learning rate of 1e-3. CosineAnnealingLR \cite{cosineannealingLR} is utilized as the scheduler with a maximum of 50 iterations and a minimum learning rate of 1e-5. Training epochs are set to 300. For VM-UNet, we initialize the weights of both the encoder and decoder with those of VMamba-S \cite{vmamba}, which is pre-trained on ImageNet-1k. All experiments are conducted on a single NVIDIA RTX A6000 GPU.

\subsection{Comparison to the State-of-the-art Models}

\subsubsection{ISIC Datasets.} We compare VM-UNet with many state-of-the-art models on the ISIC17 and ISIC18 datasets. As shown in Table \ref{tab:isic}, VM-UNet demonstrates strong performance in skin lesion segmentation tasks. Specifically, on the ISIC17 dataset, VM-UNet achieves the best results in four metrics: mIoU, DSC, Acc, and Sen. For the ISIC18 dataset, VM-UNet also outperforms other models in three metrics: mIoU, DSC, and Acc. Particularly, compared to TransFuse, a powerful Transformer-CNN hybrid model, our VM-UNet achieves a 0.72\% increase in mIoU and a 0.44\% increase in DSC.

\subsubsection{Synapse Dataset.} For the Synapse dataset, we conduct a more comprehensive comparison among CNN-based models, Transformer-based models, and hybrid models. The results are presented in Table \ref{tab:synapse}. Specifically, compared to Swin-UNet, the first pure Transformer-based image segmentation model, our VM-UNet achieves an improvement of 1.95\% and 2.34mm in terms of DSC and HD95 metrics, respectively. Particularly for the "Stomach" organ, VM-UNet achieves a DSC of 81.40\%, indicating significant gains compared to previous models. This clearly demonstrates the superiority of SSM-based models over other models in medical image segmentation tasks.

\subsection{Ablation Studies}

\begin{table}[!t]
    \centering
    \setlength\tabcolsep{4.5pt}
	\renewcommand\arraystretch{1.5}
 	\caption{Ablation studies on initial weights of VM-UNet.}
\begin{tabular}{c|cc|cc}
\hline
\multirow{2}{*}{Init. Weight} & \multicolumn{2}{c|}{ISIC17} & \multicolumn{2}{c}{ISIC18} \\ 
             & mIoU(\%)$\uparrow$         & DSC(\%)$\uparrow$         & mIoU(\%)$\uparrow$         & DSC(\%)$\uparrow$          \\ \hline
 - &77.59 &87.38 &78.66 &88.06 \\
VMamba-T     & 78.85         & 88.17       & 79.04        & 88.29        \\ 
VMamba-S     & 80.23         & 89.03       & 81.35        & 89.71        \\ \hline
\end{tabular}
    \label{tab:weight}
\end{table}

\subsubsection{Initial weights.} In this section, we conducted ablation experiments on VM-UNet initialized with VMamba-T and VMamba-S pre-trained weights on ISIC17 and ISIC18, respectively\footnote{where VMamba-T and VMamba-S achieve Top-1 accuracies of 82.2\% and 83.5\% on ImageNet-1k.}. The experimental results are presented in Table \ref{tab:weight}. Specifically, compared to VM-UNet without pre-trained weights, VM-UNet initialized with the most powerful VMamba-S demonstrates an average improvement of 2.67\% and 1.65\% in mIoU and DSC metrics across the two ISIC datasets. This reveals that stronger pre-trained weights can significantly enhance the performance of VM-UNet on downstream tasks, indicating a substantial influence of pre-trained weights on VM-UNet.

\subsubsection{Dropout Value.} To mitigate the phenomenon of overfitting in downstream image segmentation tasks, we have incorporated the Dropout technique \cite{dropout}. As illustrated in Table \ref{tab:drop}, we experimented with different Dropout values, ranging from 0.0 to 0.3. For the ISIC17 dataset, the absence of Dropout (a value of 0.0) resulted in the highest performance, with the mIoU and DSC reaching 80.45\% and 89.17\%, respectively. In contrast, for the ISIC18 dataset, the optimal mIoU and DSC were observed at a Dropout rate of 0.2, achieving 81.35\% and 89.71\%, respectively. These findings indicate a clear divergence in the impact of Dropout on the performance of the VM-UNet across different datasets. While increasing the Dropout rate appears to detrimentally affect performance on the ISIC17 dataset, a moderate level of Dropout enhances performance on the ISIC18 dataset, suggesting that the issue of overfitting might be more pronounced with the ISIC18 dataset.

\begin{table}[!t]
    \centering
    \setlength\tabcolsep{3.9pt}
	\renewcommand\arraystretch{1.5}
 	\caption{Ablation studies on the different values of dropout.}
\begin{tabular}{c|cc|cc}
\hline
\multirow{2}{*}{Dropout value} & \multicolumn{2}{c|}{ISIC17} & \multicolumn{2}{c}{ISIC18} \\ 
             & mIoU(\%)$\uparrow$         & DSC(\%)$\uparrow$         & mIoU(\%)$\uparrow$         & DSC(\%)$\uparrow$          \\ \hline
0.0     &    80.45      &   89.17     &   79.33      &     88.47    \\
0.1     &     80.02     &   88.90     &    80.23     &    89.03     \\
0.2     & 80.23         & 89.03       & 81.35        & 89.71        \\
0.3     &  80.20        & 89.02       & 79.72        &  88.71       \\ \hline
\end{tabular}
    \label{tab:drop}
\end{table}

\subsubsection{Encoder-Decoder Architectures.} 

In this paper, we design VM-UNet as an asymmetric structure, employing a strong encoder-weak decoder configuration to reduce parameter count and computational load. To elucidate the advantages of this design, we conduct a ablation study on the model architecture, the results of which are shown in Table \ref{tab:archi}. Specifically, when adopting the symmetric structure \{2,2,2,2-2,2,2,2\} for VM-UNet, not only did the parameter count increase by 0.1M, but the computational load also increased by 0.24 GFLOPs, leading to a performance decline. Furthermore, enlarging the model scale of VM-UNet to \{2,2,9,2-2,9,2,2\}, we observe a continued decline in model performance. Therefore, in this paper, we design VM-UNet as an asymmetric structure, specifically \{2,2,2,2-2,2,2,1\}.

\begin{table}[!t]
    \centering
    \setlength\tabcolsep{1pt}
	\renewcommand\arraystretch{1.0}
 	\caption{Ablation studies on different architectures of VM-UNet.}
\begin{tabular}{ccc|cc|cc}
\hline
\multicolumn{3}{c|}{Model} & \multicolumn{2}{c|}{ISIC17} & \multicolumn{2}{c}{ISIC18} \\ 
           Encoder-Decoder & \#Params & FLOPs & mIoU(\%)$\uparrow$         & DSC(\%)$\uparrow$         & mIoU(\%)$\uparrow$         & DSC(\%)$\uparrow$          \\ \hline
\{2,2,2,2-2,2,2,1\}   &27.43M &4.11G  & 80.23         & 89.03       & 81.35        & 89.71        \\
\{2,2,2,2-2,2,2,2\}   &27.53M &4.35G  &  79.23        &  88.41      & 80.74        & 89.34        \\ 
\{2,2,3,2-2,3,2,2\} &29.92M &4.81G &  78.75        & 88.11       &  80.52       &  89.21       \\ 
\{2,2,4,2-2,4,2,2\} &32.32M &5.27G  &   79.37       &   88.50     &  81.13       &    89.58     \\ 
\{2,2,9,2-2,9,2,2\}  &44.27M&7.56G  &  79.67        &   88.68     &  81.05       &   89.54      \\ \hline
\end{tabular}
    \label{tab:archi}
\end{table}

\subsubsection{Input Resolution.} 

As the length of the input sequence increases, Mamba exhibits superior performance on DNA sequence data \cite{mamba}. However, in the case of Visual Mamba, such as Vim \cite{VisionMamba} and VMamba \cite{vmamba}, a systematic ablation study regarding the length of the input sequence has not been conducted. Therefore, in this section, by increasing the resolution of input images—thereby increasing the length of the input sequence—we investigate Mamba's performance in the visual domain, especially in the medical image segmentation task. As shown in Table \ref{tab:inputsize}, despite the increase in the length of the input sequence, the performance of VM-UNet does not meet expectations. The specific reasons for this phenomenon warrant further in-depth investigation.

\begin{table}[!t]
    \centering
    \setlength\tabcolsep{3pt}
	\renewcommand\arraystretch{1.5}
 	\caption{Ablation studies on the input size of images. Due to the OOM (Out of Memory) issue of the NVIDIA RTX A6000 GPU, experiments with a resolution of 512×512 are conducted on a single NVIDIA A100 GPU.}
\begin{tabular}{c|cc|cc}
\hline
\multirow{2}{*}{Input size} & \multicolumn{2}{c|}{ISIC17} & \multicolumn{2}{c}{ISIC18} \\ 
             & mIoU(\%)$\uparrow$         & DSC(\%)$\uparrow$         & mIoU(\%)$\uparrow$         & DSC(\%)$\uparrow$          \\ \hline
256×256     & 80.23         & 89.03       & 81.35        & 89.71 \\
384×384     &  79.44        & 88.54       &  81.08       &  89.55       \\
512×512     &  70.88        &   82.96     &   72.36      &   83.96      \\ \hline
\end{tabular}
    \label{tab:inputsize}
\end{table}

\begin{figure*}[!t]
    \centering
    \includegraphics[width=0.95\textwidth]{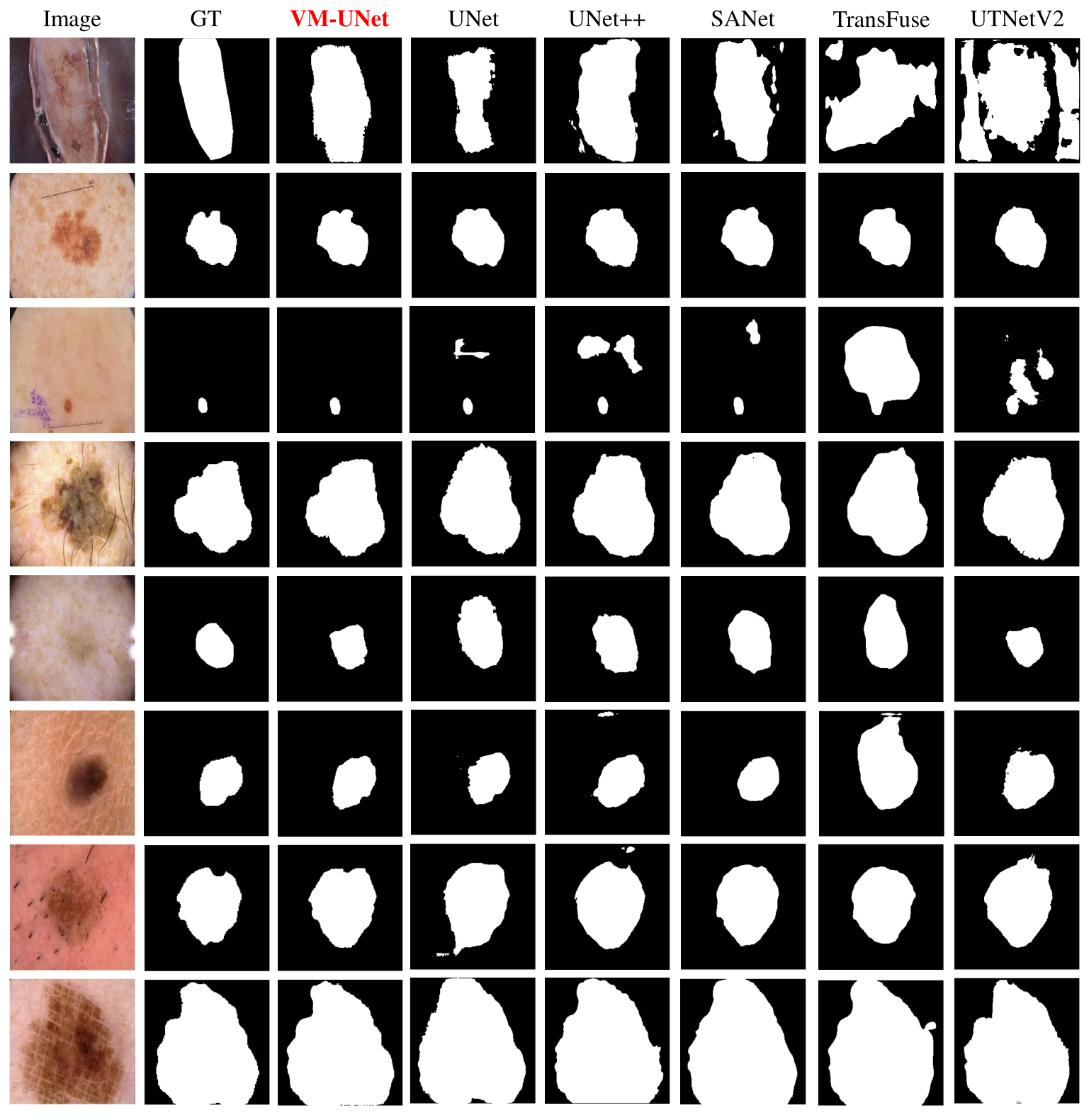}
    \caption{Visual Comparisons on the ISIC18 Dataset. GT stands for ground truth.}
    \label{fig:vis}
\end{figure*}

\subsection{Visualization}

In this section, the segmentation results of VM-UNet and other state-of-the-art models on the ISIC18 dataset are visualized, as shown in Figure \ref{fig:vis}. These results demonstrate the exceptional robustness of VM-UNet across various challenging scenarios. Specifically, in the visualizations of the first row, it is observed that non-target areas are often misclassified by other models, such as TransFuse \cite{transfuse} and UTNetV2 \cite{utnetv2}, leading to inferior segmentation performance; VM-UNet, on the other hand, exhibits greater stability in this aspect. Moreover, in the context of segmenting small targets, as illustrated in the third row of visualizations, VM-UNet outperforms other models by reducing redundant segmentation predictions. Additionally, VM-UNet is also capable of effectively handling complex segmentation boundaries, as shown in the visualizations of the fourth and last rows. Despite the complexity of the boundaries, VM-UNet still accurately delineates target edges. These visualizations further validate that VM-UNet, as a pure SSM-based model, holds significant potential in the field of medical image segmentation.

\subsection{Discussion}
In this work, we propose VM-UNet framework to explore the potential applications of purely SSM-based models in medical image segmentation. Even though our method can achieve favorable performance on some datasets, there still exists some obvious limitations in current SSM-based version:

First, structured SSMs were originally defined as discretizations of continuous systems, and have had a strong inductive bias toward continuous-time data modalities such as perceptual signals (e.g. audio, video), which may degradate the model's capability on length generalization, especially for the datasets beyond the training sequence length. Besides, the parameter count of VM-UNet is approximately 30M, providing further improvement to streamline SSMs via manual design or other compression strategies, thereby fortifying their applicability in real-world medical scenarios.

Second, the advantage of SSMs lies in their ability to capture information from long sequences, enabling a more systematic exploration of segmentation performance of pure SSM-based models at larger resolutions. However, our empirical evaluation is limited to small model sizes, below the threshold of most strong open source LLMs. These issues warrant further research and consideration when deploy our work in real-world scenarios.

\begin{figure}[!t]
    \centering
    \includegraphics[width=0.45\textwidth]{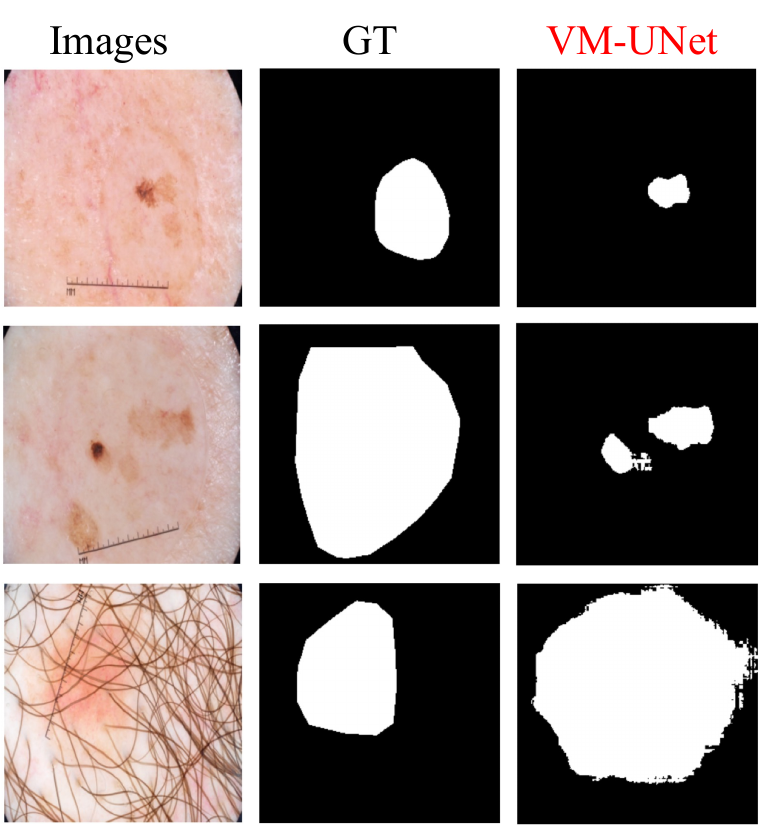}
    \caption{Some failure examples of our VM-UNnet on ISIC18 Dataset.}
    \label{fig:fail}
\end{figure}

Last but not least, the proposed method in this paper can dynamically capture extensive contextual information for medical image segmentation. Despite its promising performance on multi-organ segmentation task, we note that there are several limitations in skin lesion segmentation. For example, as shown in Figure \ref{fig:fail}, we notice that our method can only segment the dark areas and fails to produce satisfactory results for light-colored areas (1-2 row), besides, it is highly sensitive to hair disturbance (3 row) due its contextual discriminability in visual representation, which may bring negative impacts on downstream tasks. These challenges warrant further research and consideration when deploying VM-UNet model in real scenarios.

\section{Conclusions}
\label{sec5}
In this paper, we introduce a pure SSM-based model for the first time in medical image segmentation, presenting VM-UNet as a baseline. To leverage the capabilities of SSM-based models, we construct VM-UNet using VSS blocks and initialize its weights with the pretrained VMamba-S. Comprehensive experiments are conducted on skin lesion and multi-organ segmentation datasets indicate that pure SSM-based models are highly competitive in medical image segmentation tasks and merit in-depth exploration in medical community. In the future, we will continue to explore the application of SSMs in other medical imaging tasks, such as detection, registration, and reconstruction, etc.


\section*{Acknowledgments}
This work was partially supported by the National Natural Science Foundation of China under Grant No.62301315, Startup Fund for Young Faculty at SJTU (SFYF at SJTU) under Grant No.23X010501967 and Shanghai Municipal Health Commission Health Industry Clinical Research Special Project under Grant No.202340010.
The authors would like to thank the anonymous reviewers for their valuable suggestions and constructive criticisms.

\bibliographystyle{IEEEtran}


\bibliography{TCSVT}





\end{document}